\documentclass[useAMS,usenatbib,useAMS]{mn2e}

\usepackage{times}

\usepackage{graphicx}
\usepackage{amsmath, amssymb}
\usepackage{txfonts}
\usepackage{amssymb}

\usepackage{graphicx,graphics}
\usepackage{natbib}
\def\km{\,{\rm km}}
\def\cm{\,\rm cm}
\def\s{\,{\rm s}}

\def\mpc{\,{\rm Mpc}}
\def\msun{\,{\rm M_{\odot}}}
\def\erg{\,{\rm erg}}
\def\kev{\,{\rm keV}}

\def\ergs{\,{\rm erg}}

\def\yr{\,{\rm  yr}}
\def\Gpc{\,{\rm Gpc}}

\title[Research on LF and selection effect]{Research on the redshift evolution of luminosity function and selection effect of GRBs}
\author[Tan \& Wang]{Wei-Wei Tan$^{1,2}$\thanks{E-mail:
wwtan@nju.edu.cn} \& F. Y. Wang$^{1,2}$\thanks{E-mail:
fayinwang@nju.edu.cn}\\
$^1$School of Astronomy and Space Science, Nanjing University,
Nanjing 210093, China\\
$^2$Key Laboratory of Modern Astronomy and Astrophysics (Nanjing
University), Ministry of Education, Nanjing 210093, China}
\begin{document}

\date{}

\pagerange{\pageref{firstpage}--\pageref{lastpage}} \pubyear{}

\maketitle

\label{firstpage}

\begin{abstract}

We study the redshift evolution of the luminosity function (LF) and
redshift selection effect of long gamma-ray bursts (LGRBs). The
method is to fit the observed peak flux and redshift distributions,
simultaneously. To account for the complex triggering algorithm of
Swift, we use a flux triggering efficiency function. We find
evidence supporting an evolving LF, where the break luminosity
scales as $L_b\propto (1+z)^{\tau}$, with $\tau =3.5^{+0.4}_{-0.2}$
and $\tau =0.8^{+0.1}_{-0.08}$ for two kind of LGRB rate models. The
corresponding local GRB rates are $\dot{R}(0)=0.86^{+0.11}_{-0.08}
\yr^{-1}\Gpc^{-3}$ and $\dot{R}(0)= 0.54^{+0.25}_{-0.07}
\yr^{-1}\Gpc^{-3}$, respectively. Furthermore, by comparing the
redshift distribution between the observed one and our mocked one,
we find that the redshift detection efficiency of the flux triggered
GRBs decreases with redshift. Especially, a great number of GRBs
miss their redshifts in the redshift range of $1<z<2.5$, where
``redshift desert" effect may be dominated. More interestingly, our
results show that the ``redshift desert" effect is mainly introduced
by the dimmer GRBs, e.g., $P<10^{-7}\ergs /\s/\cm^2$, but has no
effect on the brighter GRBs.

\end{abstract}


\section{Introduction}

Gamma-ray bursts (GRBs) are one of the most luminous and distant
transients in the universe (Greiner et al. 2009; Cucchiara et al.
2011). Observationally, GRBs are usually categorized into two
groups: spectrally soft long GRBs with $T_{90} > 2 \s$ are expected
to result from the collapse of short lived massive stars, as some
long GRBs are evidenced to connect with Type Ic supernovae (e.g.
Galama et al. 1998; Bloom et al. 2002; Stanek et al. 2003; Thomsen
et al. 2004; Campana et al. 2006; Berger et al. 2011; Melandri et
al. 2012). Meanwhile, spectrally hard short GRBs with $T_{90} < 2
\s$ are believed to originate from the merger of compact stars (e.g.
Eichler et al. 1989; Nakar 2007; Fong et al. 2013). In this paper,
GRBs are actually referred to long GRBs, unless otherwise specified.

The hosts of GRBs in the star formation regions suggest that GRBs
could be used as the tracer of star formation rate (SFR) (Totani
1997; Paczy\'nski 1998; Wijers et al. 1998). The detection of GRB
980425 associated with a supernova also strengthened the expectation
(Galama et al. 1998). Therefore, GRBs provide a new opportunity for
the measuring of star formation history (for a recent review, see
Wang, Dai \& Liang 2015), especially at high redshift, where direct
measurement is difficult. Thanks to the launch of \emph{Swift}
satellite, which provides a large number of GRBs with measured
redshifts. This large GRB sample makes it possible to give a more
tight constraint on the relation between GRB rate and SFR, and rules
out the models that GRBs unbiased trace the cosmic star formation
history (Chary et al. 2007). Recent results show that GRBs do not
trace the star formation history exactly but with an additional
evolution, e.g., $(1+z)^\delta$ (Le \& Dermer 2007; Kistler et al.
2008; Wang \& Dai 2009; Wanderman \& Piran 2010; Cao et al. 2011;
Robertson \& Ellis 2012; Wang 2013). However, the value of $\delta$
varies large from $\delta=0.5$ to $\delta=2$ which strongly depends
on the sample selection, and the inferred SFR at high redshift seems
too high comparing with the observation from the galaxy surveys
(Kistler et al. 2008, 2009; Robertson \& Ellis 2012; Wang 2013).
Furthermore, Yu et al. (2015) even found an excess of GRB rate at
low redshift of $z<1.0$ (see also Petrosian et al. 2015). Many
theoretical models have been proposed to explain the discrepancy
between the GRB rate and SFR, e.g., the evolution of the cosmic
metallicity (Langer \& Norman 2006; Li 2008; Elliott et al. 2012),
the evolution of the initial mass function (Wang \& Dai 2011), or
the additional cosmic string explosions (Cheng et al. 2010).
However, most previous works are based on the assumption that the
luminosity function (LF) is a constant form and independent of
redshift. Alternatively, some works suggest that the LF should
evolve with redshift by using the GRB sample with pseudo redshifts
inferred from luminosity relation (e.g., luminosity-variability
relation, luminosity-peak energy relation; Lloyd-Ronning et al.
2002; Firmani et al 2004; Yonetoku et al. 2004; Matsubayashi et al.
2005; Kocevski \& Liang 2006; Tan et al. 2013). The evolution of the
LF scales as $L\propto (1+z)^\tau$, with $\tau$ varies from 1 to 3.
In this paper, we will use GRBs with measured redshifts, but not the
pseudo ones, to find out whether LF evolves with redshift or not.

In this work, we specially pay attention to two important effects.
One is the redshift selection effect. Previous works use the GRB
redshift distribution to study the LF or GRB rate do not consider
the missing redshift problem (e.g. Kocevski \& Liang 2006;
Salvaterra et al 2012). However, this selection effect is very
important for these studies, as only $\sim 30\%$ of \emph{Swift}
GRBs have measured redshifts. Especially, if the selection effect
evolves with redshift, the results even could be wrong. A detailed
study about the redshift selection effect is presented in Coward et
al. (2013) (see also Fiore et al. 2007). The other effect is the
\emph{Swift} triggering threshold problem. Before \emph{Swift}, a
single detection threshold based on an increased photon count rate
above the background is a reasonable approximation. However,
\emph{Swift} has a much more complex triggering algorithm in order
to maximize the GRB detection (Band 2006), e.g., the \emph{Burst
Alert Telescope} (BAT) on board \emph{Swift} has over 500 rate
trigger criteria. Therefore, a single detection threshold
approximation may not be correct for \emph{Swift}. To solve this
problem, we use a more complex triggering algorithm which is based
on the most recent result of Lien et al. (2014).

First, we study the redshift evolution of the LF by fitting the
number distributions of peak flux and redshift. Considering the
discrepancy between the SFR and GRB rate, we introduce two kind of
GRB rate models. While fitting the redshift distribution, only
redshifts measured by absorbtion spectroscopy are considered (to
reduce the redshift selection effect; see the following section for
details). For the peak flux data, we take it from the Butler's
online catalog
\footnote{http://butler.lab.asu.edu/Swift/bat\_spec\_table.html}.
Then, we analyze the redshift selection effect by comparing the
number distribution between the observed redshift sample (include
all GRBs with measured redshifts) and the model predicted one. We
find that the selection effect is not only redshift dependent but
also flux dependent.

This paper is organized as follows. In section 2, we describe the
data extraction methodology. In section 3, models applied to
constrain the LF and redshift selection effect are presented.
Results are presented in section 4. The summary and discussion will
be given in section 5.

\section{The data}

\begin{figure}
\centering\resizebox{0.4\textwidth}{!}{\includegraphics{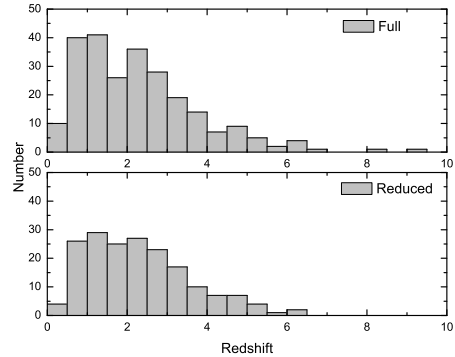}}
\caption{Top: The redshift distribution of the full GRB sample
including 244 GRBs. Bottom: The reduced redshift distribution of 182
GRBs with redshifts obtained either from only afterglows, or from
both afterglows and host galaxies.}\label{figure 1}
\end{figure}

Including long and short GRBs, \emph{Swift} has detected more than
900 GRBs (until GRB 141026A) in the past 10 years. However, recent
studies suggest that the classical classification of short and long
GRBs with $T_{90}=2\s$ is not appropriate for \emph{Swift}. Bromberg
et al. (2013) argued that a more suitable selection for long GRBs
from \emph{Swift} satellite should be defined by $T_{90}>0.8\s$,
which is based on a physically motivated model. Here we believe that
the classification with $T_{90}=2\s$ is more strict in our study. As
shown in Figure 3 of Bromberg et al. (2013), the probability of a
GRB with $T_{90}>2\s$ to be a collapsar is more than $80\%$, which
is higher than $50\%$ for $T_{90}>0.8\s$. Additionally, three
sub-luminous GRBs (GRB 060218, 060505, 100316D) with luminosity
$L<10^{49} \erg\s^{-1}$ have been excluded from our sample
(Soderberg et al. 2006; Liang et al. 2007; Virgilii et al. 2008;
Howell et al. 2011; Howell \& Coward 2013), three ultra-long GRBs
(GRB 101225A, 111209A, 121027A) have been suggested to be another
population(Gendre et al. 2013; Stratta et al. 2013; Levan et al.
2014). We also exclude 22 possible short GRBs (GRB 050416A, 050724,
050911, 051016B, 051114, 051227, 060505, 060614, 061006, 061210,
070714B, 071227, 080123, 080503, 080520, 090531B, 090607, 090715B,
090916, 100213A, 100216A, 100816A; Dietz 2011; Kopa\v{c} et al.
2012; Fong et al. 2013; Tsutsui et al. 2013; Howell \& Coward 2013;
Berger 2014; Howell et al. 2014).

In this paper, the redshift data was taken from the \emph{Swift}
archive \footnote{http://swift.gsfc.nasa.gov/docs/swift/archive/grb
table.}. After removing the uncertain redshifts, we obtained 282
GRBs (including long and short GRBs) with measured redshifts, which
is around 30\% of the BAT-triggered GRBs. For long GRBs, their
redshifts are always measured from afterglows and/or the host
galaxies observed by the ground-based telescopes, including the Very
Large Telescope (VLT), Gemini-S-N, Keck and Lick. Finally, we
obtained 244 long GRBs with confirmed redshifts (the full redshift
sample). However, this full redshift sample is greatly affected by
the observation biases. On one hand, more low redshift GRBs could be
observed, as the brightest GRBs are predominantly nearby. On the
other hand, some redshift dependent selection effects even could
change the shape of the redshift distribution. As shown in the upper
panel of Figure 1, a large number of GRBs missed their redshifts in
the range of $1.5<z<2$, which may caused by the ``redshift desert"
effect. Therefore, to derive a less biased redshift sample is an
immediate necessity of this work. Following Fynbo et al. (2009), we
reconstruct our redshift sample by considering redshifts measured
either from afterglows, or from both afterglows and host galaxies.
We exclude the bursts with redshifts measured from the emission
spectra of the host galaxies, as these bursts always cover a low
redshift range (e.g., $0.3<z<2.8$; Wanderman \& Piran 2010, Howell
et al. 2014). We also exclude GRBs with photometric redshifts
because of the large uncertainties. After doing this, we obtained a
less biased redshift sample with 182 GRBs (the reduced redshift
sample). The distribution of the reduced redshift sample is shown in
the bottom panel of Figure 1. It's obvious that the completeness of
the reduced redshift sample is better than that of the full redshift
sample, as shown in the top and bottom panels of figure 1 (e.g., in
the redshift range of $1.5<z<2$). However, one may argue that the
missing redshifts of GRBs in the range of $1.5<z<2$ may be caused by
the statistical fluctuations. For this point, we will discuss it in
section 4.1.

In this work, we use the GRB peak flux data in the energy range of
15-150$\kev$ of the BAT. The data was taken from the Butler's online
catalogue, which is the extension of the work presented in Butler et
al. (2007, 2010). As the peak flux data can be measured directly
from BAT by assuming the fewest burst characters, we suggest that
the peak flux distribution is the least uncertain property. Finally,
we obtained a total number of 681 bursts with peak fluxes.

\section{The model}

\begin{figure}
\centering\resizebox{0.4\textwidth}{!}{\includegraphics{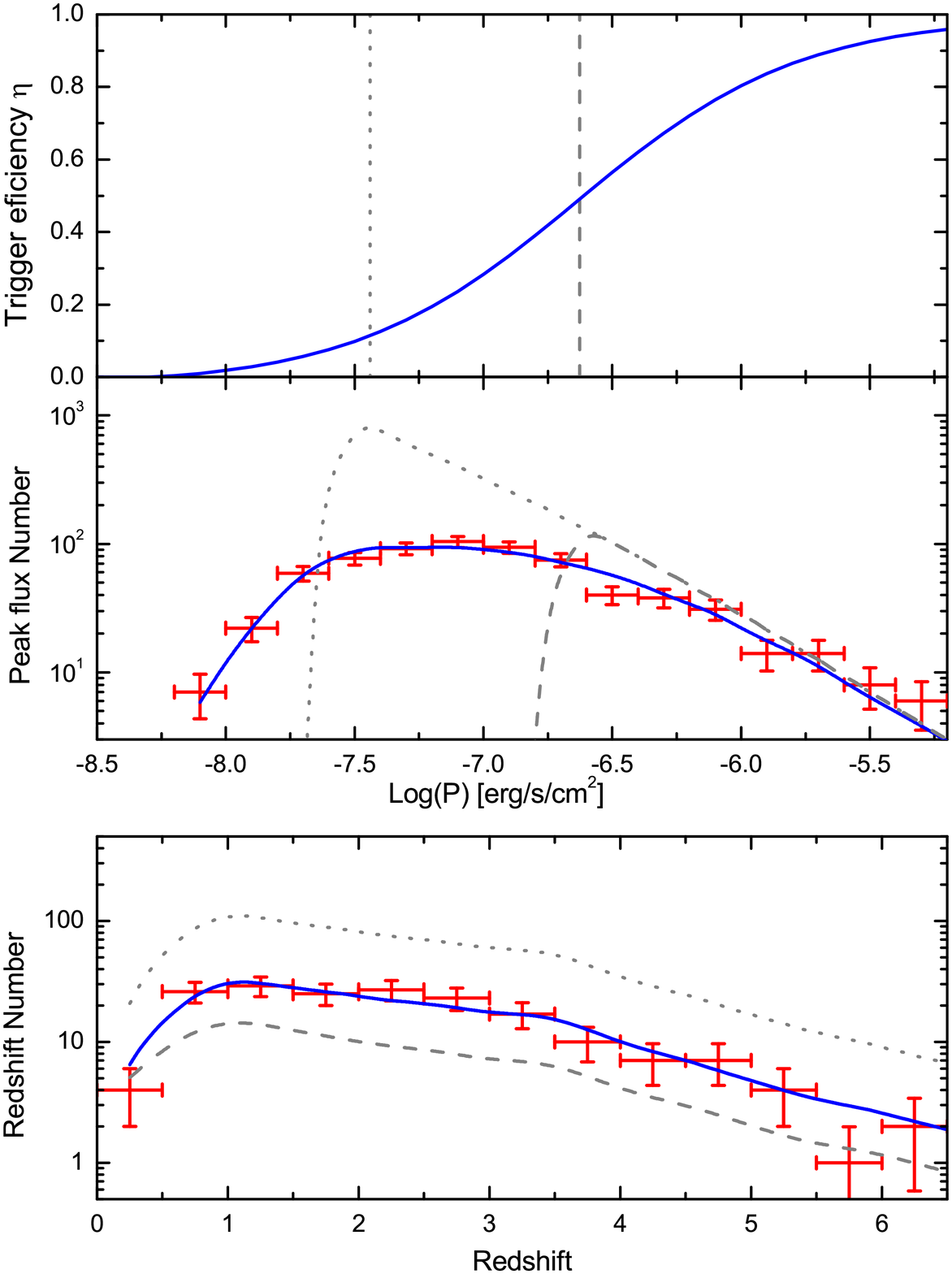}}
\caption{Top: The triggering efficiency of \emph{Swift} as a
function of peak flux (blue solid line), and the single detection
threshold of 2.6 ph $\s^{-1}\cm^{-2}$ (gray dashed line) and 0.4 ph
$\s^{-1}\cm^{-2}$ (gray dotted line). The middle and bottom figures
show the effects of the triggering algorithm on the number
distributions of peak flux and redshift (for RGRB1 for example),
respectively.}\label{figure 2}
\end{figure}

The first step of our work is to fit the GRB number distributions of
peak flux and redshift, which are both related to the GRB rate and
LF. The expected number of GRBs with observed peak fluxes between
$P_1$ and $P_2$ that triggered the BAT can be expressed by
\begin{eqnarray}
N(P_1, P_2)={\Delta\Omega\over 4\pi} T
\int^{\infty}_{0}\int^{L_{2}}_{L_{1}}\eta(P)\Phi_{z,P}(L)\dot{R}(z)dL{dV(z)\over
1+z}, \label{expected P}
\end{eqnarray}
where $(\Delta\Omega/4\pi)\sim0.1$ is the field view of the BAT,
$T\sim10 ~\rm yrs$ is the observational period, $dV(z)$ is the
comoving volume and $1/(1+z)$ accounts for the time dilation,
$\eta(P)$ is the triggering function, $\Phi_{z,P}(L)$ is the LF with
$L_{\min}=10^{49}\erg\s^{-1}$ and $L_{\max}=10^{57}\erg\s^{-1}$ for
normalization, and $\dot{R}(z)$ is the GRB rate. The expected number
of GRBs within redshift range of $z_1<z<z_2$ is given by
\begin{eqnarray}
N(z_1, z_2)={\Delta\Omega\over 4\pi} T
\int^{z_2}_{z_1}\int^{L_{\max}}_{L_{\rm
min}}\eta(P)\Phi_{z,P}(L)\dot{R}(z)dL{dV(z)\over 1+z}.
\label{expected z}
\end{eqnarray}
For the flat $\Lambda$CDM cosmology, we employ the cosmological
parameters from WMAP nine-year results with $\Omega_m=0.28,
\Omega_\Lambda=0.72$ and $H_0=70 \km\s^{-1}\mpc^{-1}$.

The triggering algorithm of \emph{Swift} is complicated. Lien et al.
(2014) simulated 50,000 GRBs to mock the triggering efficiency,
where GRB light curves, incident angles, BAT's active detector
number, and the image trigger algorithm are considered. Although a
specific set of parameters are included, we consider that the
triggering algorithm is better than other techniques. By comparing
the simulated peak flux number distribution with the real triggered
one, Howell et al. (2014) derived the flux triggering efficiency in
a functional form of
\begin{eqnarray}
\eta(P)={a(b+c P/P_0)\over (1+P/d P_0)} \label{trigger function},
\end{eqnarray}
with $P>5.5\times 10^{-9}\erg\s^{-1}\cm^{-2}$, and below this range
the function equals to zero. The function parameters are as follows:
a=0.47, b=-0.05, c=1.46, d=1.45 and $P_0=1.6\times
10^{-7}\erg\s^{-1}\cm^{-2}$ (Howell et al. 2014). We could reproduce
the triggered GRB population in Lien et al (2014) by using this
triggering function. It seems that the sample completeness is much
better than that of the step function approximation. In figure 2, we
compare $\eta(P)$ with two step function approximations, e.g., a
value of 0.4 ph $\s^{-1}\cm^{-2}$ and a value of 2.6 ph
$\s^{-1}\cm^{-2}$, which have been used by Salvaterra et al. (2012).
From the middle panel of the figure, we see that the threshold of
0.4 ph $\s^{-1}\cm^{-2}$ predicts more dim GRBs, while the threshold
of 2.6 ph $\s^{-1}\cm^{-2}$ predicts more bright ones. It is
apparent that the adoption of such approximations could be
problematic while estimating the relative contributions of bright
and dim GRBs. Furthermore, it seems that the adoption of different
triggering algorithms do not change the shape of the redshift
distribution significantly, but change the GRB number greatly as
shown in the bottom panel of the figure. So it is obvious that the
threshold of 0.4 ph $\s^{-1}\cm^{-2}$ predicts a lower local GRB
rate, while the threshold of 2.6 ph $\s^{-1}\cm^{-2}$ predicts a
higher one.

\begin{figure*}
\resizebox{0.8\hsize}{!}{\includegraphics{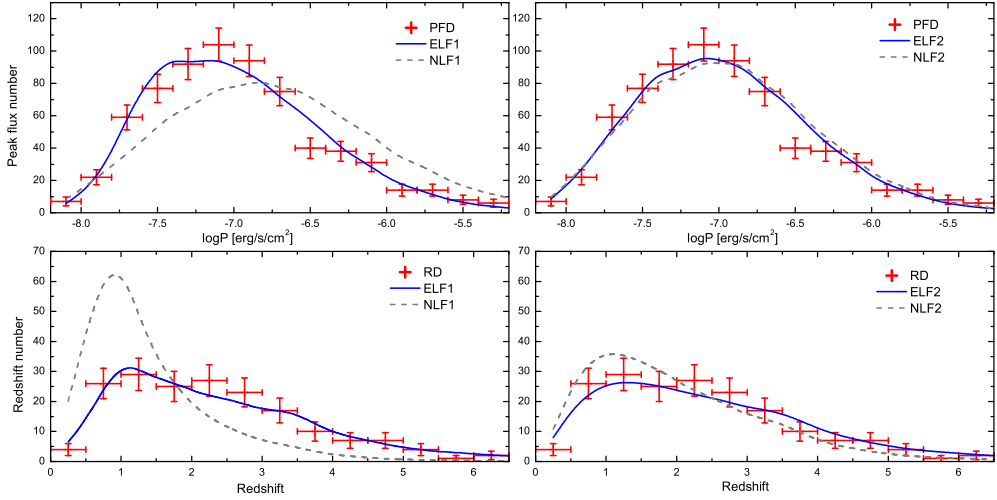}}
\caption{Left: The number of peak flux distribution (top; PFD) and
redshift distribution (bottom; RD) of RGRB1. The solid blue lines
are our best fitting results with an evolving LF (ELF1). The dashed
gray lines are the fitting results with an non-evolving LF (NLF1).
Right: The same as the left figures but for RGRB2.}\label{fig3}
\end{figure*}

The isotropic-equivalent peak luminosity in the source frame can be
calculated by
\begin{eqnarray}
L=4\pi d_{\rm L}(z)^2 P {k(z)\over b},
\end{eqnarray}
where $P$ is the peak flux (in units of $\erg\s^{-1}\cm^{-2}$) in
the observed energy band of $15-150 \kev$, $d_{\rm L}(z)=(1+z)
d_c(z)$ is the luminosity distance. The term ${k(z)/ b}$ is to
convert the observed energy band to the rest frame band of $1-10^4
\kev$. Here $b$ accounts for the bolometric energy fraction that
seen in the detector band (Wanderman \& Piran 2010, Howell et al.
2011), and $k(z)$ is the cosmological correction. We express them in
the form of
\begin{eqnarray}
b=\int^{150}_{15}E S(E)dE/\int_{1}^{10^4}E S(E)dE
\end{eqnarray}
and
\begin{eqnarray}
k(z)=\int^{150}_{15}E S(E)dE/\int_{15(1+z)}^{150(1+z)}E S(E)dE,
\end{eqnarray}
respectively. Here $S(E)$ is the rest frame photon spectrum, which
can be well expressed by the empirical Band function (Band et al.
1993, 2003). The high and low energy spectral indices are given by
-2.25 and -1, respectively. The spectral peak energy in the source
frame can be derived by the Yonetoku relation (Yonetoku et al. 2004)
\begin{eqnarray}
E_p(1+z)=\left({2.34\times10^{47}\over L}\right)^{-1/2},
\end{eqnarray}
which is a much tighter and reliable relation.

We assume a broken power law form for the LF, which is suggested to
be better than that of a single power law(e.g., Cao et al. 2011; Tan
et al. 2013)
\begin{eqnarray}
\Phi_{z,P}(L)\propto\left\{~\begin{array}{ll}\left({L\over
L_b(z)}\right)^{-\alpha},~~~~& L\leq L_b(z),\,\\
\left({L\over
L_b(z)}\right)^{-\beta},~~~~&L>L_{b}(z),\,\end{array}\right.
\label{luminosity function},
\end{eqnarray}
where the break luminosity is assumed to evolve with redshift in the
form of $L_b(z)=\mathcal{C}\times(1+z)^{\tau}$. The normalization
coefficient is taken by assuming the minimum luminosity of $L_{\rm
min}=10^{49}\erg\s^{-1}$. Here $\alpha$, $\beta$, $\tau$ and
$\mathcal{C}$ are the free parameters which will be determinated in
the following section.

Finally, we consider two kind of GRB rate models. For the first one,
we consider that GRB rate follows the SFR (RGRB1 for short). The
observational GRB production rate can be connected to star formation
rate as
\begin{eqnarray}
\dot{R}_1(z)=f_bf_c\dot{\rho}_{*}(z),
\end{eqnarray}
where $f_b=0.01$ is the beaming degree of GRB outflows and $f_c$
arises from the particularities of GRB progenitors (e.g., mass,
metallicity, magnetic field, etc). We assume $f_bf_c$ does not
evolve with redshift, and the redshift evolution effect between GRB
rate and SFR mainly originates from the LF.

For the SFR $\dot{\rho}_{*}(z)$ , we describe it as (Hopkins \&
Beacom 2006)
\begin{eqnarray}
\dot{\rho}_*(z)\propto \left\{~~\begin{array}{ll}(1+z)^{3.44},&
~~~~z\leq0.97,\,\\
(1+z)^{0}, & ~~~~0.97<z\leq3.5,\,\\
(1+z)^{\kappa}, & ~~~~3.5<z,\,
\end{array}\right.\label{star formation rate}
\end{eqnarray}
with the local star formation rate $\dot{\rho}_*(0)=0.02\rm M_\odot
yr^{-1}Mpc^{-3}$. Here we assume the high-redshift SFR ($z>3.5$)
evolves in the form of a power law, which is still ambiguous.
However, motivated by the shapes of the SFR at $z<3.5$ and also
implied by some preliminary measurements (Bouwens et al. 2008, 2011;
Ellis et al. 2012) and GRBs as well as their host galaxies (Y\"uksel
et al. 2008; Kistler et al. 2009; Wang \& Dai 2011; Elliott et al.
2012), the power low form of the SFR is considerable. For
simplicity, we set a typical value of $\kappa=-3$ in the following
calculation (e.g., Wang 2013).

The second GRB rate model (RGRB2 for short) is derived from the
observed long GRBs. Following Wanderman \& Piran (2010), we describe
it in a functional form of
\begin{eqnarray}
\dot{R}_2(z)=\dot{R}_0 \left\{~~\begin{array}{ll}(1+z)^{v_1},&
~~~~z\leq z_*,\,\\
(1+z)^{v_2}, & ~~~~z_*<z,\,\\
\end{array}\right.\label{GRB rate2}
\end{eqnarray}
with typical values of $z_*=3.6, v_1=2.1$, and $v_2=-0.7$ based on
the recent study (e.g., Howell et al. 2014). This form of GRB rate
model is widely used in literature.

\section{Results}
In this section, we constrain the model parameters by jointly
fitting the peak flux distribution (PFD) and redshift distribution
(RD; the reduced redshift sample). As mention in the above section,
two GRB rate models are considered. In each model, five free
parameters are to be constrained: $\alpha$, $\beta$, $\tau$,
$\mathcal{C}$, and $f_c$ (for RGRB1) or $\dot{R}_0$ (for RGRB2). In
fact, $f_c$ or $\dot{R}_0$ can be easily derived by equating the
model predicted peak flux number (equation 1) to the observed one
(number of 681), if other four parameters are constrained. Other
three parameters of $\alpha$, $\beta$ and $\mathcal{C}$ are mainly
determinated by the PFD. The lower P-values of PFD fix the values of
$\alpha$ and $\mathcal{C}$, and $\beta$ is mainly determined by the
higher P-values. For the value of $\tau$, it is strongly dependent
on the redshift evolution of RD.

\begin{table*}
\begin{center}
\begin{tabular}{c|cccccccccccc}
\hline \hline
Model&$\alpha$&$\beta$& $\tau$& $\mathcal{C}$& $f_c$&$\dot{R}(0)$& $\chi^2$ &$Q$& $\chi_{\rm PFD}^2$  &$Q_{\rm PFD}$  &$\chi_{\rm RD}^2$ &$Q_{\rm RD}$  \\
       &          &    &  & $10^{50}\erg\s^{-1}$ &  $10^{-6}\msun^{-1}$ &$\yr^{-1}\Gpc^{-3}$&&&&&&\\
\hline
ELF1     &$0.3^{+0.11}_{-0.2}$~ & $1.9^{+0.05}_{-0.05}$~ & $3.5^{+0.4}_{-0.2}$ ~ & $0.33^{+0.1}_{-0.055}$~ & $4.34^{+0.6}_{-0.39}$~ & $0.86^{+0.11}_{-0.08}$&17.74&0.77& 12.31  &0.26&5.43&0.71\\

ELF2     &$0.1^{+0.73}_{-0.1}$~ & $2.1^{+0.13}_{-0.03}$~ & $0.8^{+0.1}_{-0.08}$ ~ & $14^{+6}_{-6}$~ & $--$~ & $0.54^{+0.25}_{-0.07}$&23.63&0.42&16.43&0.09&7.20&0.51\\

NLF1     &$0.1^{+0.1}_{-0.1}$~ & $1.7^{+0.05}_{-0.05}$~ & $0$ ~ & $3^{+0.5}_{-0.3}$~ & $6.9^{+0.2}_{-0.2}$~ & $1.36^{+0.04}_{-0.04}$&296.77&$\sim$0&197.33&$\sim$0&99.44&$\sim$0\\
NLF2     &$0.1^{+0.13}_{-0.1}$~ & $2.4^{+0.12}_{-0.14}$~ & $0$ ~ & $39^{+5.6}_{-3.2}$~ & $--$~ & $0.66^{+0.06}_{-0.05}$&39.2&0.026&21.45&0.029&17.75&0.038\\
\hline
\end{tabular}
\end{center}
\caption{The best-fit parameters for an evolving LF (ELF) and
non-evolving LF (NLF) for RGRB1 and RGRB2 (with number 1 and 2
represent the two models), respectively. The statistical quality
parameters of the models are shown in the last six
columns.}\label{tbl-1}

\end{table*}

Before the start of the work, we first set a primary constraint on
the related parameters, e.g., $\alpha>0$, $\beta>0$,
$L_b(0)>10^{49}\erg\s^{-1}$, $\tau>0$, and the local GRB rate
$\dot{R}(0)\in[0,2]\yr^{-1}\Gpc^{-3}$ for both GRB rate models.
Then, we give an arbitrary set of values for the five free
parameters in the allowed ranges. For each set of the parameters, we
calculate the $\chi^2$ values for both PFD ($\chi^2_{\rm PFD}$) and
RD ($\chi^2_{\rm RD}$). The total $\chi^2$ is assumed to be the
linear combination of $\chi^2_{\rm PFD}$ and $\chi^2_{\rm RD}$. The
best fit parameters are derived by minimizing the global $\chi^2$.
At last, we obtain the error bars for the five parameters by setting
$\chi^2=\chi^2_{\min}+5.89$, which corresponds to the parameters
within 68.3\% confidence level. Table 1 gives the values of the
fitted parameters for each model. The degrees of freedom (for the
simultaneous fit) in Table 1 is 28 minus the number of parameters.
The information about the quality of the fit for models is also
given in the table. For each model, we show the total $\chi^2$ and
the goodness-of-fit \textit{Q} (the probability to find a new
$\chi^2$ exceeding the current one).

In figure 3, we plot the observed and model-predicted PFD and RD.
The left and right figures are the fitting results for RGRB1 and
RGRB2, respectively. The solid lines are for models with the
evolving LFs, while the dashed lines are for models without
evolution. From this figure, we conclude our results as follows:

1. For both GRB rate models, the evolving LFs (ELF) fit well with
both PFD and RD, while the non-evolving LFs (NLF) fit the data
poorly, which can be inferred from the values of the $\chi^2$ and
\textit{Q} in Table 1. For RGRB1, the non-evolving LF (NLF1) model
shows an excess of GRBs at high \textit{P} and low redshift (left
panels in figure 3). For RGRB2 (NLF2), it also shows the same
tendency as RGRB1 (right panels in figure 3). Therefore, we suggest
that the LF should evolve with redshift and rule out the
non-evolving LF for both GRB rate models.

2. A non-negligible fraction of high-redshift GRBs may exist in the
current \emph{Swift} GRB sample. In our reduced redshift sample, the
highest redshift is $z=6.32$ (GRB 140515A). By using the best fit
parameters provided in Table 1, we find nearly 23 GRBs with their
redshifts higher than 6.5 for RGRB1. Simulatively, nearly 24 GRBs
with redshifts higher than 6.5 for RGRB2. However, the truth is that
we only observed 4 GRBs with possible redshifts higher than that
(GRB 060116, $z=4$ or 6.6; 080913, $z=6.44$ or $6.7$; 090423,
$z=8.26$; 090429B,$z=9.3$). This result suggests that a large
fraction of high-redshift GRBs (even with redshift higher than 10;
e.g., Tan et al. 2015) may exist in the BAT-triggered GRB sample.
Unfortunately, we can not measure their redshifts because of the
instrumental reasons. It is also possible that these high-redshift
GRBs may origin from Pop III stars (e.g., M\'{e}sz\'{a}ros \& Rees
2010; Toma et al. 2011).

3. Except for the redshfit evolution of LF, an additional evolution
may exist between the GRB rate and SFR. If the LF and GRB rate model
of RGRB2 represent the intrinsic ones, the additional redshift
evolution is obvious while comparing the values of $\tau$ between
the two models ($\tau=3.5^{+0.4}_{-0.2}$ for RGRB1 and
$\tau=0.8^{+0.1}_{-0.08}$ for RGRB2). However, it is hard to
describe this evolution form as it may originate from many reasons,
e.g., the evolution of cosmic metallicity (Wang \& Dai 2009; Li
2008), initial mass function, different properties of progenitor,
their host galaxies, or their combination. The collapsar model
explains the formation of GRB via the collapse of a rapidly rotation
massive iron core into a black hole (Woosley 1993). However, there
are two basic problems of single-star model in producing collapsars.
First, the spin-down of the stellar core due to core-envelope
coupling. Second, the Wolf-Rayet winds can slow the rotation.
Fortunately, the rapid rotation of low-metallicity stars can
overcome both problems (Langer \& Norman 2006). Such speculation is
confirmed by observation (Fynbo et al. 2003; Modjaz et al. 2008;
Graham \& Fruchter 2013; Trenti et al. 2013; Wang \& Dai 2014). So
we believe that it is easier for low-metallicity stars to produce
GRBs than the stars with high-metallicity. If we incorporate this
effect into the LF (as we have done in the model of RGRB1, where
only the redshift evolution of LF is considered, but ignore the
evolution of the GRB progenitor properties), it will lead to a
stronger redshift evolution of the LF, as the cosmic metallicity
decreases with redshift. Futhermore, much more massive stars exist
at high redshift, which will lead to more energetic GRBs and also
increase the LF evolution. Therefore, the intrinsic redshift
evolution of the LF should in the range of
$0.8^{+0.1}_{-0.08}<\tau<3.5^{+0.4}_{-0.2}$ for reality.

\subsection{Redshift selection effect}
\begin{figure}
\centering\resizebox{0.45\textwidth}{!}{\includegraphics{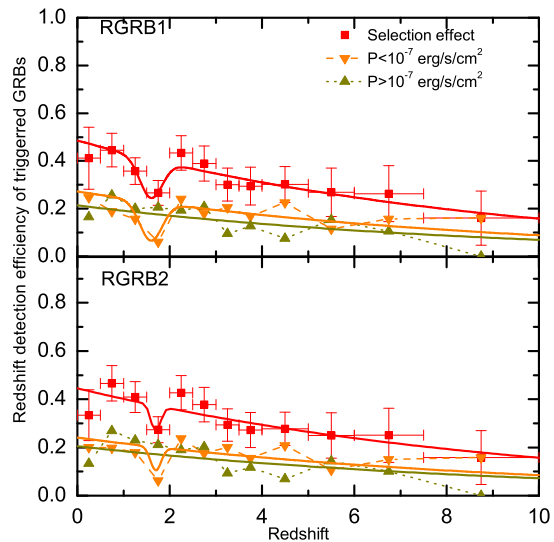}}
\caption{Redshift detection efficiency of BAT-triggered GRBs of
\emph{Swift}, with a total number of 244 GRBs with redshifts
measured from afterglows and/or host galaxies. Top: Redshift
detection efficiency for RGRB1. The solid squares correspond to the
detection efficiency of the total GRBs. The up-triangles
(down-triangles) with dotted (dashed) line correspond to the
detection efficiency of GRBs with peak fluxes
$>10^{-7}\erg\s^{-1}\cm^{-2}$ ($<10^{-7}\erg\s^{-1}\cm^{-2}$). The
solid lines are our fitting results. Bottom: The same as that of the
top panel, but for RGRB2.}\label{figure 4}
\end{figure}

There are two major elements to affect the GRB redshift detection
efficiency: one is the instrumental bias and the other is the
redshift-dependent selection effect. For the first point, the
localization and sensitivity of the follow-up afterglow observations
have great impact on the detection probability of GRB redshifts. The
quick response and high sensitivity making \emph{Swift} detect the
largest number and the highest redshift of GRBs. Assuming GRB
optical afterglows decrease like power laws with exponent
$\gamma\sim-1$, \emph{Swift} will detect GRB afterglows with 2.9 and
4.0 mag brighter than $HETE2$ and $BeppoSAX$ (Fiore et al. 2007),
respectively. For the second point, the host galaxy dust obscuration
plays a most important role in the measurement of GRB redshifts. It
is suggested that a fraction of 30-35 percent of GRBs miss their
redshifts for dust obscuration (Coward et al. 2013). In addition,
the so-called ``redshift desert" effect can also lead to a fraction
of nearly $ 20\%$ of GRBs missing their redshifts in the range of
$1.1<z<2.1$ (Steidel et al. 2005; Fiore et al. 2007; Coward et al.
2013). The Malmquist bias arises because of the sensitivity of
telescopes and instruments, which leads to a decreasing redshift
detection efficiency to higher redshift. Therefore, we only have
$\sim 30\%$ of GRBs with measured redshifts.

In Figure 4, we compare the number distribution of the full redshift
sample with our model predicted one in each redshift bin. The ratio
could be considered as the redshift detection efficiency (solid
squares). The error bars along the y-axis are the statistical errors
(i.e., square root of the number in each bin $\Delta N=\sqrt{N}$),
which correspond to the 68\% Poisson confidence intervals for the
binned events (Gehrels 1986). The x-axis error bars simply represent
the bin size. For both GRB rate models, the detection efficiency
have the tendency of decreasing with redshift. Particularly, nearly
20\% of GRBs miss their redshifts in the range of $ 1<z<2.5$, which
could be the evidence of the ``redshift desert" effect. Furthermore,
we divide the full redshift sample into two groups: one with peak
fluxes $P>10^{-7}\ergs /\s/\cm^2$ (up-triangles with dotted line),
and the other with $P<10^{-7}\ergs /\s/\cm^2$ (down-triangles with
dashed line). It is obvious that the ``redshift desert" effect
mainly results from the low peak flux GRBs with $P<10^{-7}\ergs
/\s/\cm^2$. For GRBs with $P>10^{-7}\ergs /\s/\cm^2$, the detection
efficiency just decreases with redshift and has no relation with the
``redshift desert" effect. See from the figure, we could find more
high (or less low) redshift GRBs at low (high) peak fluxes, which
may also indicate that high (low) redshift GRBs always have low
(high) peak fluxes. For simplicity, we first fit the detection
efficiency of RGRB1 with an exponential function multiple a Gauss
function as
\begin{eqnarray}
\eta_{\rm t}(z)=0.36\times e^{0.3 - z/8.9}\times(1 - 0.41\times
e^{-(z - 1.6)^2/0.11}).
\end{eqnarray}
Then we fit the sample with $P>10^{-7}\ergs /\s/\cm^2$ by
multiplying an constant with equation (12) but without the Gauss
fraction, and the fitting result could be expressed as
\begin{eqnarray}
\eta_{\rm h}(z)=0.16\times e^{0.3 - z/8.9}.
\end{eqnarray}
For the detection efficiency of $P<10^{-7}\ergs /\s/\cm^2$, we
obtain it by subtracting equation (13) from (12), which is
$\eta_{\rm l}(z)=\eta_{\rm t}(z)-\eta_{\rm h}(z)$. It seems that
$\eta_{\rm l}(z)$ also fits the reuslt quite well, which could also
support our result that the ``redshift desert" effect mainly comes
from low flux GRBs. For RGRB2 model, we go through the same steps
and the fitting results are as follows,
\begin{eqnarray}
\eta_{\rm t}'(z)=0.33\times e^{0.3 - z/9.6}\times(1 - 0.3\times
e^{-(z - 1.7)^2/0.02})
\end{eqnarray}
and
\begin{eqnarray}
\eta_{\rm h}'(z)=0.15\times e^{0.3 - z/9.6}.
\end{eqnarray}
Finally, we obtain $\eta_{\rm l}'(z)=\eta_{\rm t}'(z)-\eta_{\rm
h}'(z)$, which also fits the observation quit well. The fitting
results are shown as the solid lines in Figure 4. The goodness of
the fits at the highest redshift are greatly affected by the GRB
number. Here we should mention that the ``redshift desert" effect
exists in both GRB rate models, particularly in the dimmer GRB
sample. However, if this is caused by the statistical fluctuations,
the ``redshift desert" effect will disappear. Then, the redshift
detection efficiency will be nearly a constant at $z<2.5$ and
decrease at higher redshift. The problem is that the fluctuations
only exist in the dimmer GRBs, but not in both dimmer and brighter
GRB samples. The so-called ``redshift desert" is a redshift region
($1<z<2.1$) where it is difficult to measure absorption and emission
spectra of the sources. As redshift increases beyond $z \sim 1$, the
strong emission lines from the host galaxies (e.g. the [O
III]$\lambda\lambda$4959, 5007, [O II]$\lambda$3727 line, H$\alpha$,
H$\beta$) are shifted outside of the typical interval covered by
optical spectrometers ($\approx 3800 - 8000 {\AA}$) at $z>1$, while
Lyman-$\alpha$ enters the range at $z\sim 2.1$. This becomes more
serious for faint sources, as the spectra of the faint sources
always have small signal to noise ratio. Therefore, we believe that
dimmer GRBs are more responsible for the ``redshift desert" effect,
which can not be caused by the statistical fluctuations.

\section{Summary and discussion}

In this work, we utilized both the peak flux and redshift
distributions to constrain the LF. We don't use the pseudo redshifts
inferred from the empirical luminosity relations, because these
relations could be greatly affected by the observational bias, e.g.,
trigger efficiency (Shahmoradi \& Nemiroff 2011). By jointly fitting
the PFD and RD, we found that the non-evolving LF fits the data
poorly, while the evolving LF fits data well. The evolution of the
break luminosity can be expressed as $L_b\propto (1+z)^\tau$, with
$\tau=3.5^{+0.4}_{-0.2}$ and $\tau=0.8^{+0.1}_{-0.08}$ for RGRB1 and
RGRB2, respectively. The strong and weak redshift evolution of the
LFs for two GRB rate models might imply an additional redshift
evolution between GRB rate and SFR. The strong redshift evolution of
LF for RGRB1 is based on the assumption that only LF evolves with
redshift, which is not real if GRB progenitors evolve with redshift.
If we take these effects into account, the redshift evolution of the
LF may become weaker for RGRB1 (e.g., $\tau <3.5^{+0.4}_{-0.2}$).
However, it is impossible to distinguish this additional evolution
under the current situation, unless the number of GRBs with measured
redshifts is large enough to satisfy the LF fittings in different
redshift ranges (Tan et al. 2013).

The local GRB rates are $\dot{R}(0)_1=0.86^{+0.11}_{-0.08}
\yr^{-1}\Gpc^{-3}$ and $\dot{R}(0)_2=
0.54^{+0.25}_{-0.07}\yr^{-1}\Gpc^{-3}$ for RGRB1 and RGRB2,
respectively, which are consistent with the previous works (e.g.,
Schmidt 1999, 2001; Guetta et al. 2004, 2005; Liang et al. 2007;
Wanderman \& Piran 2010; Cao et al. 2011; Lien et al. 2014).
Furthermore, we suggest that a large fraction of high-redshift GRBs
($z>6.5$) may have been detected by \emph{Swift} satellite but
without measured redshifts. These high-redshift GRBs may be produced
by Pop III stars (e.g., M\'{e}sz\'{a}ros \& Rees 2010; Toma et al.
2011; Tan et al. 2015).

To reduce the large biases introduced by the complex triggering
criterion of \emph{Swift}, we use the recent peak flux efficiency
function for the triggering criterion instead of the constant photon
flux (Howell et al. 2014). The constant photon flux triggering
criterion could be problematic while estimating the contribution of
the dim and bright GRBs as shown in figure 2. The introduction of a
dependence of $L$ on $E_p$ has little effect on our results, and the
Yonetoku relation is considered to be tight enough.

To reduce the redshift selection effect, only GRBs with redshifts
measured from afterglows or from both afterglows and host galaxies
are considered. After deriving the best-fit parameters, we compare
the observed full redshift distribution with our model predicted
one. Our results show that the redshift detection efficiency
decreases with redshift slowly in general, and nearly 20\% of GRBs
missed their redshifts in the ``redshift desert" range of $
1<z<2.5$. The reason for the slow decreasing tendency of the
detection efficiency might be the combined effects of the dust
extinction and Malmquist bias, as the higher (lower) redshift GRBs
always have less (more) dust contents but lower (higher) peak fluxes
(e.g. Bouwens et al. 2009; Perley et al. 2009; Zafar et al. 2011;
Rossi et al. 2012). By dividing the GRB sample into the dim and
bright groups, we find that the ``redshift desert" effect is mainly
introduced by the low-flux GRBs, which is a useful result for the
completeness of the sample selection in future works.

\section*{Acknowledgements}
We greatly acknowledge the referee, Prof. Robert J. Nemiroff, for
the valuable comments, which have significantly improved our work.
We also thank H. Yu for the technical help. This work is supported
by the National Basic Research Program of China (973 Program, grant
No. 2014CB845800) and the National Natural Science Foundation of
China (grants 11422325, 11373022, 11103007, and 11033002), the
Excellent Youth Foundation of Jiangsu Province (BK20140016).

\end{document}